\documentclass[fleqn,usenatbib]{mnras}

\usepackage{newtxtext,newtxmath}

\usepackage[T1]{fontenc}

\DeclareRobustCommand{\VAN}[3]{#2}
\let\VANthebibliography\thebibliography
\def\thebibliography{\DeclareRobustCommand{\VAN}[3]{##3}\VANthebibliography}

\usepackage{graphicx}	
\usepackage{amsmath}	
\usepackage{subcaption}
\usepackage{bm}
\usepackage{booktabs,siunitx,threeparttable}
\usepackage{adjustbox,makecell}

\newcommand\PSmatched{\texttt{PS1-matched}}
\newcommand\Gaiamatched{\texttt{Gaia-matched}}

\newcommand{\mean}{\langle m \rangle}
\newcommand{\xmean}{\langle x \rangle}
\newcommand{\median}{\mathrm{Median}}

\usepackage{booktabs}
\usepackage{array}  
\usepackage{longtable}
\graphicspath{{Images/}}

\title[M-dwarf flare catalogue]{The largest ground-based catalogue of M-dwarf flares}

\author[A. D. Lavrukhina et al.]{
A.~D.~Lavrukhina,$^{1,2}$\thanks{E-mail: lavrukhina.ad@gmail.com}
B.~Demkov,$^{3}$
K.~Malanchev,$^{4}$
M.~V.~Pruzhinskaya$^{1}$ and
E.~E.~O.~Ishida$^{5}$
\\
$^{1}$Lomonosov Moscow State University, Sternberg astronomical institute, Universitetsky pr.~13, Moscow, 119234, Russia\\
$^{2}$Lomonosov Moscow State University, Faculty of Physics, Leninskie Gory 1-2, Moscow, 119991, Russia\\
$^{3}$Independent Researcher\\
$^{4}$McWilliams Center for Cosmology \& Astrophysics, Department of Physics, Carnegie Mellon University, Pittsburgh, PA 15213, USA\\
$^5$Université Clermont Auvergne, CNRS, LPCA, Clermont-Ferrand, F-63000, France}

\date{Accepted XXX. Received YYY; in original form ZZZ}

\pubyear{\the\year{}}

\begin{document}
\label{firstpage}
\pagerange{\pageref{firstpage}--\pageref{lastpage}}
\maketitle

\begin{abstract}
We present the largest ground-based catalogue of M-dwarf flares to date, comprising 1,229 time-resolved events identified from the Zwicky Transient Facility (ZTF) data release 17.  
Using high-cadence ZTF observations collected between April 2018 and September 2020, we analyzed over 93 million variable light curves containing 4.1 billion photometric measurements.  
Flare candidates were initially identified through a machine-learning pipeline trained on simulated flare light curves, generated by injecting TESS-based flare templates into ZTF observational data.  
The candidates were then processed through an extensive post-filtering pipeline combining a machine-learning classifier, additional metadata gathering, and human inspection.  
For 655 flares with reliable Gaia-based distance estimates and well-sampled light-curve profiles, we derived bolometric energies ranging from $10^{31}$ to $10^{35}$ erg.
A clear correlation is observed between flare frequency and spectral subtype, with a sharp increase toward later M~dwarfs, particularly near M4-M5, coinciding with the transition to full convection.
Using the subset of 680 flaring stars with known vertical distances from the Galactic plane $z$, we estimated how the fraction of flaring stars varies with Galactic height and found a negative trend, indicating that the likelihood of flare activity decreases with increasing $|z|$.
The resulting catalogue represents the most comprehensive ground-based sample of M-dwarf flares available and establishes a framework for flare detection and classification in upcoming wide-field surveys such as the Vera C. Rubin Observatory Legacy Survey of Space and Time.
\end{abstract}




\section{Introduction}

M-dwarf flares are highly energetic transient phenomena characterized by rapid energy release, with typical bolometric energies ranging from $\sim10^{29}$ to $10^{36}$ erg and durations spanning from minutes to several hours~\citep{tess_dr1_flares, Hawley_2014}. 
These events manifest across the entire electromagnetic spectrum, from radio to X-ray wavelengths, with the most dramatic enhancements typically observed in the UV. 
The physical origin of these flares is intimately linked to magnetic reconnection processes and dynamo mechanisms~\citep{Hawley_1991, Kowalski_2013}.
The enhanced magnetic activity observed in M- dwarfs stems from their fully convective internal structure, which facilitates efficient magnetic field generation and amplification~\citep{Browning_sym_conv_stars, Kitchatinov}.
The study of M-dwarf flares consequently represents a critical intersection of stellar and planetary science, as frequent energetic outbursts may significantly impact the habitability of nearby exoplanets.

There is a widely supported hypothesis that young stars exhibit flares more frequently than the older ones~\citep{West_rotation, Newton_rotation}.
This can be attributed to the stronger magnetic fields generated by the rapid rotation of stars, which is characteristic of young stellar objects.
The idea that younger stars produce flares more often has strong backing from both observational data and theoretical understanding.
However, in order to confirm this trend with higher statistical confidence, larger and more homogeneous flare samples across a broad range of stellar ages are still required.

The abilities of modern time-domain surveys has enabled systematic searches for stellar flares across a wide range of spectral types, with M dwarfs being of particular interest due to their strong magnetic activity. 
Space-based missions such as \textit{Kepler} \citep{2010Sci...327..977B} and the Transiting Exoplanet Survey Satellite (TESS; \citealt{2014SPIE.9143E..20R}) provide continuous monitoring with high photometric precision, allowing the detection of both frequent small-amplitude flares and rare energetic events. 
For example, \citet{yang2019_kepler_flares} reported $\sim1.6\times10^5$ flares on about 3400 stars from \textit{Kepler} data, while \citet{yang2019_kepler_flares} identified 8695 flares in the first TESS data release and \citet{tess_3year_flares} extended this number to $\sim1.4\times10^5$ flares over the first three years of the mission. 
CHEOPS high-cadence monitoring combined with wavelet-based denoising enabled the detection of 291 microflares with energies as low as $10^{26}$ erg \citep{poyatos2025_flares}. 
More recently, \citet{galletta2025_flares} presented a TESS-based catalog of 17,229 flares from nearby M dwarfs ($d\lesssim10$ pc), probing energies down to $10^{29}$ erg and demonstrating the large diversity of flare occurrence rates among stars of similar luminosity.

Ground-based surveys complement the space missions by covering larger sky areas and probing brighter, rarer flares. 
The Zwicky Transient Facility (ZTF; \citealt{2019PASP..131a8002B}) has been employed for systematic flare searches, including machine-learning approaches based on active anomaly detection, which identified more than one hundred flares \citep{Voloshina_2024}. 
Other surveys have also contributed valuable flare samples: hundreds of flares in Sloan Digital Sky Survey (SDSS; \citealt{SDSS}) and Two Micron All Sky Survey (2MASS; \citealt{2MASS}) data with maximum $u$-band amplitudes up to $\sim4.5$ mag \citep{sdss_2mass_flares}, 62 flares up to $\sim2$ mag in $V$-band in the ASAS-SN M-dwarf catalog \citep{asas-sn_flares}, 96 flares from DECam data \citep{DECam_flares}, 22 fast flares from the Tomo-e Gozen project with on
e-second cadence \citep{tomoe-gozen_flares}, and 132 flares in the TMTS survey \citep{tmts_flares}.

The remaining of this paper is organized as follows. Section~\ref{sec:data} describes the ZTF data release (DR) 17 dataset, the high-cadence subset, and all preprocessing steps. 
Section~\ref{sec:pipeline} details the flare-search pipeline, including  cadence and variability-based pre-filtering (Section \ref{sec:pre-filtering}), \ref{sec:flare-simulation} simulating proccess of flares built from TESS templates for training, \ref{sec:light-curve-classification} feature extraction and classification with Random Forest and CatBoost combined via a blending ensemble, and \ref{sec:post-filtering} post-filtering using Minor Planet Checker\footnote{\url{https://www.minorplanetcenter.net/cgi-bin/checkmp.cgi}} (MPChecker), image-quality metrics, and Pan-STARRS1 (PS1) colors. 
Section~\ref{sec:results} presents the resulting catalog of M-dwarf flares, statistics across spectral subtypes and Galactic height,  energy -- duration estimates and their correlations. 
Finally, the Appendix provide the full list of expert-crafted features.

\section{Data}\label{sec:data}

The Zwicky Transient Facility (ZTF) is a wide-field sky survey conducted with the 48-inch Schmidt-type Samuel Oschin Telescope at Palomar Observatory~\citep{2019PASP..131a8002B}. 
It has a field of view of 47 square degrees observed in the $g$, $r$, and $i$ passbands, with a typical 30-second exposure reaching a median $r$-magnitude limit of approximately 20.6~mag.
In phase I (March 2018 – Sept 2020), ZTF observed the northern sky with a three-day cadence and the Galactic plane with a one-day cadence. In phase II (Oct 2020 – Sept 2023), half of the ZTF telescope time is dedicated to a uniform two-day cadence public survey in the $g$- and $r$- bands. Data from this public survey are released every two months as data releases. 

The ZTF DR photometric pipeline performs source extraction for individual frames and then cross-matches these sources across all frames within a single pair of observation field/CCD quadrant and passband. This process leads to a single sky source potentially being represented by multiple ZTF DR objects due to overlapping observation fields and the three passbands used.

In this work we use public and private survey data from ZTF DR17 as target data sets for searching for M-dwarf flares. 
This release covers the time span from March 2018 to March 2023.
The DR~17 source catalog files contain 756 billion source detections, and 4.66 billion light curves derived by cross-matching the single-exposure data to the ZTF reference catalog.

M-dwarf flares are rapid events, requiring a high observational cadence to detect and accurately capture their flare profiles. The specialized high-cadence Galactic plane survey, conducted by \citealt{Kupfer_2021} with a 40-second cadence, is particularly well-suited for this purpose. 
All observations of the high-cadence Galactic plane survey were obtained in $r$-band.
Unfortunately, the bulk-downloadable data releases do not include any metadata indicating the specific survey to which individual observations belong. As a result, we implemented a specialized filtering process based on the number of observations, the time delay between consecutive observations, and the total duration of the corresponding segment of the light curve (for further details, see Section~\ref{sec:pre-filtering}).
The outcome of this process is a dataset of light curve segments that satisfies the filtering criteria. Since multiple segments from the same light curve may meet these criteria, it is possible for more than one segment to share the same ZTF object identifier (OID). To differentiate between them, each light curve segment is characterized by both its ZTF OID and the first observation's heliocentric modified Julian date (HMJD).

\section{Search pipeline}\label{sec:pipeline}

Usually, M-dwarf flare light curves have single- or multi-modal fast-rise exponential-decay profiles.
These profiles can be detected with numerical fitting, especially in the cases of high signal-to-noise ratio and good observational cadence.

Even in these cases some false positive detections are still possible. For example, if time series are not long enough to also include some observations before and after the flare.
In such cases, other types of variability may produce profiles which would have a satisfactory model fit, for example, a single period of a fast-pulsating star (see Section~\ref{sec:visual-inspection}).
This type of false positives may be reduced by using a non-parametric light-curve shape model; in this paper, we use machine learning algorithms.
Machine learning models may also help to detect flares with unusual shapes, reducing false-negative rate if examples of such shapes are given in the training set.

Despite these improvements, in the output of our machine learning models, we found many types of light curves causing false positive detections: image defocusing issues, asteroid occultations, etc.
Such problems can be solved best by fetching additional data from external services, which is not feasible for the whole input data set.
Even after all machine learning and metadata filters, expert analysis of candidate flares is required.
This leads us to a "funnel" pipeline where each step reduces the number of candidates to an  amount feasible for the next step.

Here is the outline of this pipeline:
\begin{itemize}
\item Section~\ref{sec:pre-filtering}: ZTF light-curve data are prefiltered based on cadence and variability.
\item Section~\ref{sec:flare-simulation} and Section~\ref{sec:light-curve-classification}: classification model selects candidates based on their light curves.
\item Section~\ref{sec:post-filtering}: filtering of selected candidates based on the image quality, external data sources, possible asteroid occultations, and stellar colours.
\item Section~\ref{sec:visual-inspection}: expert analysis of the candidates.
\end{itemize}

We made use of computational cloud resources for data pre-processing, training, prediction, and post-filtering.
A computational node with 32 CPU cores and 256 GB of RAM was used for the following tasks: data pre-processing took 19 hours, model training took 4 hours, and prediction took 26 hours.

\subsection{Pre-filtering}\label{sec:pre-filtering}
The first step in the search pipeline is data selection, which is analogous to what we have done in \cite{Voloshina_2024}.
Since we are specifically looking for stellar flares, our focus is limited to variable light curves.
To form a dataset of variable light curves, we used a reduced $\chi^2$ statistics for the constant magnitude model (number of parameters $d=1$):
\begin{equation}
\label{reduced_chi2}
    \text{reduced}~\chi^2 \equiv \frac1{N-d} \sum_i\left(\frac{m_i - \bar{m}}{\delta_i}\right)^2,
\end{equation}
where $m_i$ and $\delta_i$ are magnitude and its observational error, $\bar{m}$ is the  weighted mean magnitude, and $N$ is the number of observations. 
The pre-filtered dataset consists of light curves with a value of the reduced $\chi^2$ statistics greater than 3.

The second goal of pre-filtering was to extract observations made during the high-cadence campaign.
In the absence of image-processing pipeline meta-information in the bulk-downloadable ZTF DRs, we selected the light curves for our final sample based on the following criteria:
$r$-band observations only; at least 10 observations per light curve; at most 30 minute delay between two subsequent observations; total duration is at least 30 minutes.

The final number of light curves after application of all described criteria is 93,660,131, the total number of detections is 4,145,625,951, the mean number of observations per light curve is $\sim 44$.
The data spans from April 2018 to September 2020, which indicates a probable ZTF observation strategy change in ZTF Phase II.
It is important to mark that in the final sample we can have several light curves which relates to the same object, since during our procedure we segmentate all such light curves to the pieces which satisfy described criteria.
The total number of unique ZTF DR object identifiers is 54,800,806, but the actual number of sources may be a little lower due to observational field overlaps.

\subsection{Flare Simulation}\label{sec:flare-simulation}
\begin{figure}
\center{
\includegraphics[width=0.5\textwidth]{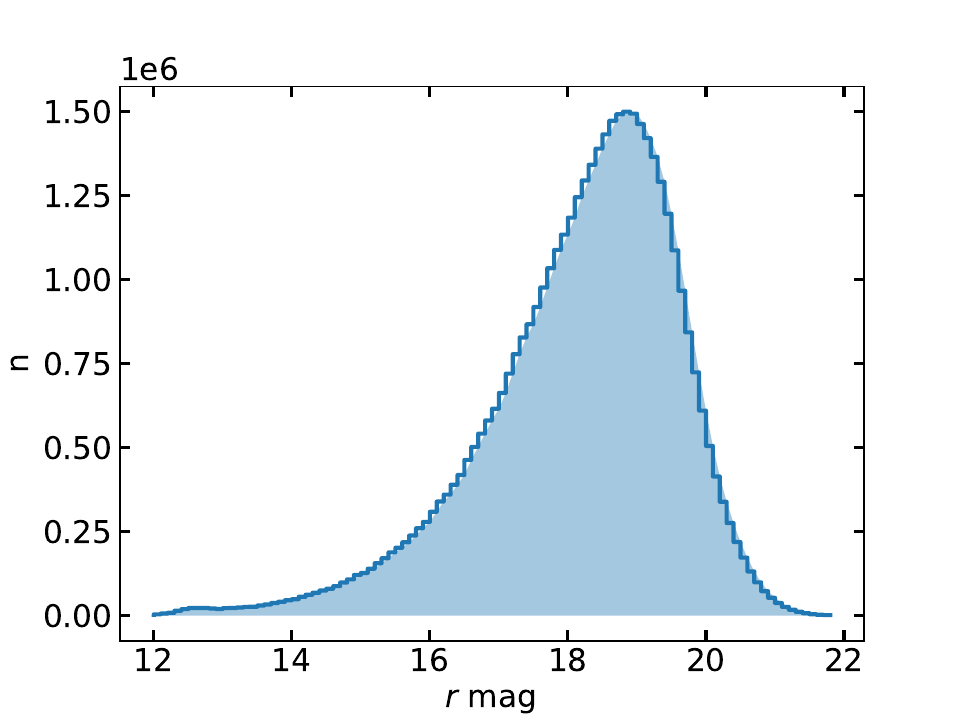}
}
\caption{Distribution of magnitudes in \textit{r}-passband from ZTF DR17 high-cadence data.}
\label{fig:mag-distr}
\vspace{0.4cm}
\end{figure}

The absence of a comprehensive sample of M-dwarf flares, detected by  ZTF, and sufficient for training a ML model, leads to the necessity of flare simulations. 
In this section, we describe a process of flares simulation based on M-dwarf flares detected in the TESS survey data~\citep{Davenport_2014}. 
The proposed method of flares simulation is based on utilizing  TESS survey data as a starting point and combining it with time grids and observational magnitudes and errors from ZTF data.
We prefiltered TESS light curves based on relative amplitude and intrinsic noise.
A total of 436 unique TESS light curves were utilized as templates. Multiple flares were detected in some of these curves, resulting in 769 extracted flare segments (templates) that were subsequently used for further simulation.
The full pipeline of flares simulation consists of further steps:
\begin{enumerate}
    \item Original light curves from the TESS survey are in fluxes, so as a first step for the initial template light curve all observations and observational errors were transformed from fluxes to magnitude.
    The magnitude light curve is linearly interpolated to produce a magnitude template function of time, $f(t)$.
    \item For each simulated flare, we randomly select the ZTF observational field from the distribution based on all high-cadence light curves.
    \item We randomly select a time-series from all ZTF light curves belonging to high cadence observations. Here and below, the high-cadence observations are selected by the same criteria as listed in Sect.~\ref{sec:pre-filtering}, but without $\chi^2$-based filtering.
    \item The peak time $t_0$ of the simulated flare is selected  
    randomly between the beginning and the end of the time series.
    This allows us to simulate different situations of flare detection (e.g. a full light curve, only descending or only ascending part) and is used partially as a data augmentation method together with selecting different time grids.
    \item In order to make the stellar magnitude distribution realistic, we sampled the quiescent magnitude $m_0$ from the empirical distribution presented in ZTF high-cadence $r$-band observations, see Fig.~\ref{fig:mag-distr}. 
    The result ``perfect'' (noise-free) magnitude is a sum of the quiescent magnitude and the time-dependent template magnitude $f(t-t_0)$.
    \item The next step involved generating the observational magnitude uncertainty.
    We assume that the noise amplitude is distributed conditionally over the source magnitude and observational conditions, which both vary over time.
    For this reason we sample noise amplitude $\sigma_i$ from an empirical distribution of ZTF errors, 
    measured on the same exposure (ZTF \texttt{expid}) and CCD quadrant (\texttt{fieldid} \& \texttt{rcid}) as the simulated observation.
    This empirical distribution is also conditional on magnitude: we use linear grid with {50 bins} between maximal and minimal value of observed mag to bin observations.
    \item As a final step, we simulate observational magnitude as a Gaussian random variable based on the ``perfect'' magnitude and the noise amplitude:
    $$m_i \sim N(f(t_i - t_0) + m_0,\, \sigma_i^2),$$
    where $m_0$ is the magnitude of the quiescent star, $m_i$ is the observed magnitude, $f(x)$ is the linear interpolation of the TESS template, $t_i$ is the time of the observation, $t_0$ is the shift due to the selection of peak time,  and $\sigma_i$ is the observation error.\footnote{Our exploration of the ZTF data showed that Gaussian distribution with provided observational error describes instrumental noise pretty well for $m\gtrsim15$~mag. We tested it with the empirical distribution of the relative residual from the mean object magnitude $(m_i - \bar{m})/\sigma_i$. The distribution was found to be sufficiently close to the standard normal distribution.}
\end{enumerate}

\begin{figure*}
    \centering
    \includegraphics[width=\textwidth]{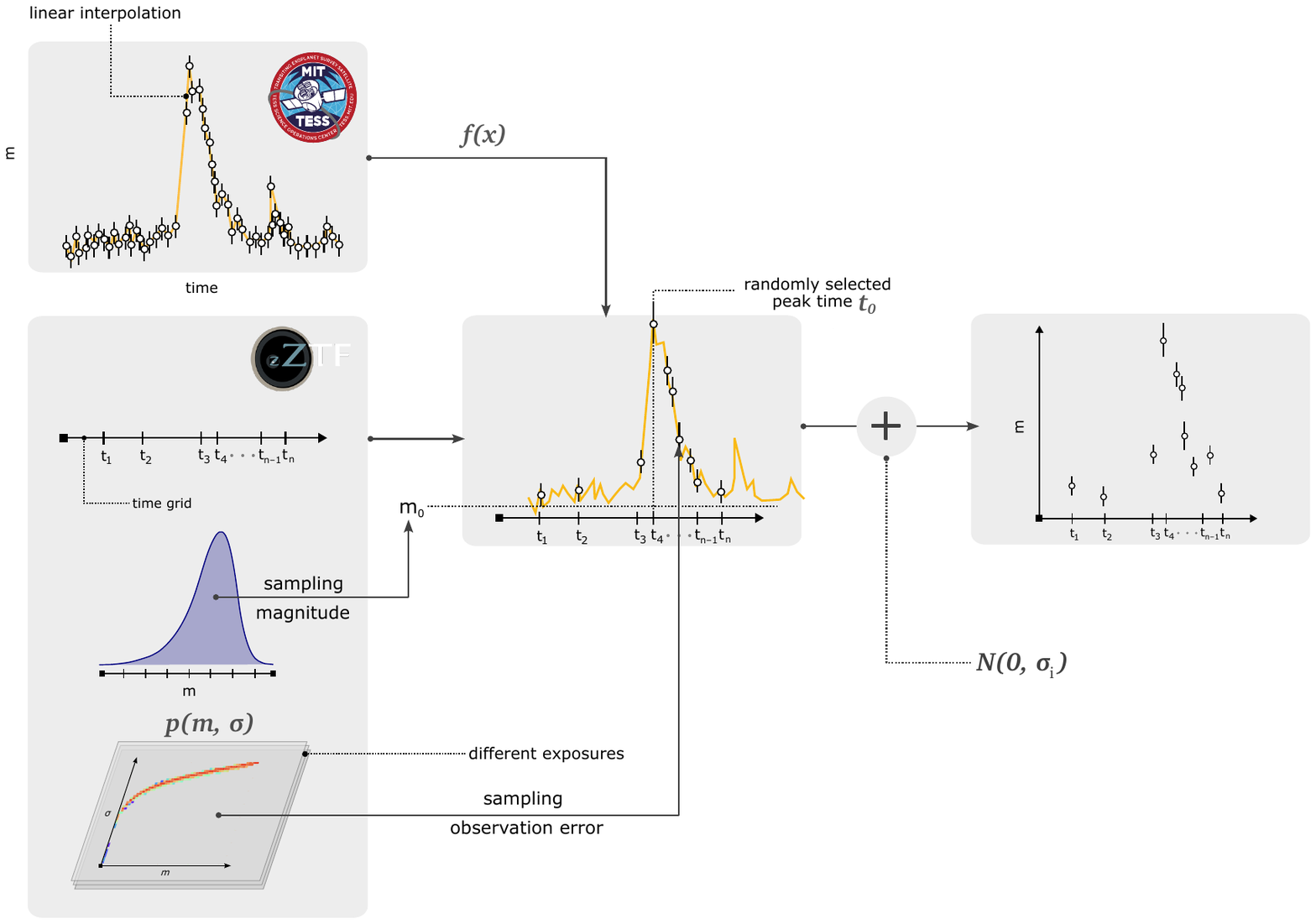}
    \caption{Flowchart of the flare light-curve simulation process. Original TESS flare templates are converted from flux to magnitude, interpolated, and injected into randomly selected ZTF high-cadence time grids with realistic noise and brightness distributions. Further details of the simulation procedure are provided in Section~\ref{sec:flare-simulation}.}
    \label{fig:simulation-flowchart}
\end{figure*}

The simulation pipeline is sketched as a flow chart in Fig.~\ref{fig:simulation-flowchart}.

\textbf{}

\subsection{Light-curve classification}\label{sec:light-curve-classification}

We developed a machine learning model for selecting flaring candidates from the pre-filtered dataset.
For model training, we generated 620\,755 flaring curves by the procedure described in the previous section; these light curves have ``positive'' labels.
For the ``negative'' class we randomly selected 620\,755 from $\sim94\times10^6$ objects of the target dataset described in Section~\ref{sec:pre-filtering}, assuming that the chance of accidentally including a flare light curve is negligible.
The result labeled dataset was split into training and test datasets in 3:1 ratio.

Since our simulation pipeline may produce some light-curve features unique to artificial light curves, the machine-learning models we train may become biased toward using these features for binary classification instead of, or in addition to, those describing stellar flares.
If this occurs, it could lead to less effective classification, as the models would not encounter these artificial features in real data and would assign lower scores to real flares.
To address this, we used an additional test set consisting of 99 ZTF light curves identified by ~\citet{Voloshina_2024}.
After training and validation of the models using the datasets described above, we perform a final quality test on this small dataset to make the final decision on model selection.

Next, in this section, we provide a detailed examination of the feature extraction process, light-curve classification models used, and metrics applied.

\subsubsection{Feature extraction}

Since number of detections, duration, and cadence of light curves are very diverse over the sample, we pre-process them before applying machine learning analysis.
We extract two different classes of features: 33 expert-selected features, and 9,996 \texttt{MiniRocket} features.
Expert-selected features are applied to the original light curves using \texttt{light-curve}\footnote{\url{https://github.com/light-curve/light-curve-python}} package, the inputs are arrays of time, magnitude and magnitude errors~\citep{light_curve}.
The full list of expert-selected features is given by Appendix~\ref{sec:features}.

The second feature set is obtained using \texttt{MiniRocket} convolution algorithm~\citep{Dempster_2021}.
First, light curves are interpolated with a third-order Akima spline to a uniform time grid of 100 points, spanning from the minimum to the maximum observational time, using the \texttt{scipy} package \citep{2020SciPy-NMeth}.
The interpolated light curves are normalized with the standard scaler: magnitude values are scaled and shifted so the mean is zero and the variance is unity.
\texttt{MiniRocket} employs a set of 84 predefined convolution kernels of size nine, with kernel values of either 2 or $-1$.
The interpolated data is convolved with these kernels, and the resulting features are quartiles of the convolution outputs.

To reduce the number of highly correlated \texttt{MiniRocket} features and balance their contribution with expert-selected features, we apply principal component analysis (PCA) to reduce the dimensionality from 9,996 to 47.
The number of PCA components is chosen such that 90\% of the total data variance is explained by the PCA model.

As the result, the total of 80 features are extracted for each light curve and machine learning models are applied to them.

\subsubsection{Classification models}

The input feature vectors are high-dimensional and may have highly correlated components due to the residual correlation of PCA-transformed \texttt{MiniRocket} features and the correlation between some of the expert-selected features~\citep[see Figure A3 of][]{Malanchev_2021}.
This motivated us to use decision-tree ensemble models for the machine-learning classification pipeline, as they are typically more robust to feature correlation.
Another advantage of decision-tree models is that their hyperparameters can be more easily configured to prevent overfitting, which was a significant risk in this research due to the use of both simulated and real data in the training dataset.

We trained two models using the same training dataset: Random Forest~\citep{random_forest} and CatBoost~\citep{dorogush2018}.
We used the \texttt{scikit-learn}~\citep{scikit-learn} implementation of Random Forest, setting the following parameters to non-default values: the number of trees to 100 and the tree depth to 36 (the last was selected by the grid-based hyperparameter tuning).
For CatBoost, we set the learning rate to 0.01, the tree depth to 10, and the maximum number of iterations to 5000. Additionally, we monitored the learning curve progress every 10 iterations for an early termination.
We set 99:1 true-to-false class weights for both algorithms.
Both models output scores are calibrated to much probability of flare class in the validation set.

We composed the models to create a third ensemble model, Blender.
Blender aggregates outputs from both Random Forest and CatBoost and yields the minimum calibrated score.
This approach makes the model more strict to false positives: for a given threshold score the object is classified as a flare, only if both models classified it so.

\begin{figure}
    \centering
    \includegraphics[width=0.45\textwidth]
    {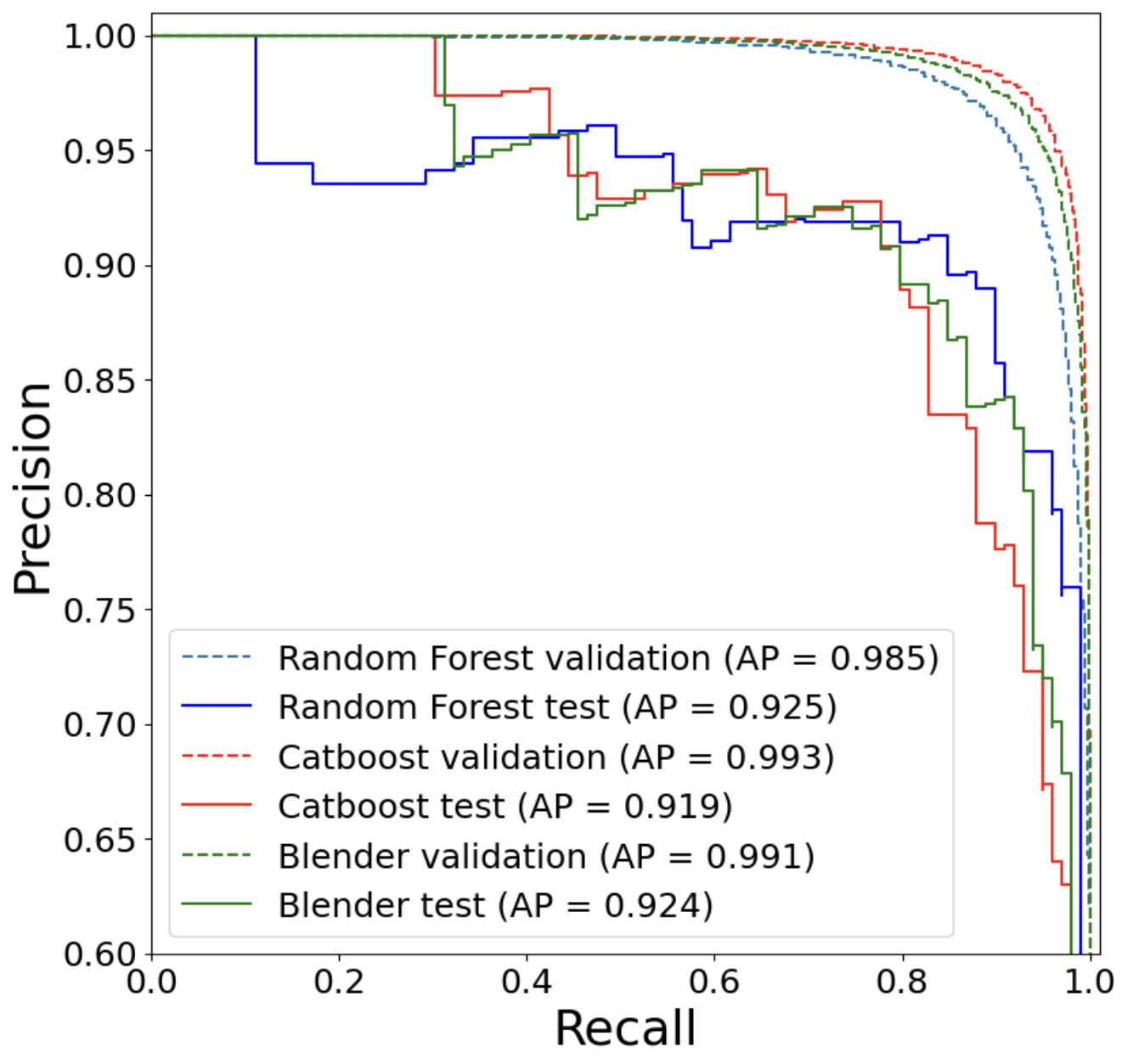}
    \caption{Precision Recall curve for validation and test datasets.}
    \label{fig:pr_curve}
\end{figure}

Table~\ref{tab:test_metrisc} shows the performance of the models according to the following metrics:
\begin{itemize}
  \item precision: ratio of correctly labeled flares to all objects labeled as flares for the default classification score threshold value;
  \item recall: ratio of the correctly labeled flares to all flares for the default threshold value;
  \item F1 score is a harmonic mean of the precision and the recall;
  \item average precision score (AP): integral under the precision--recall curve, see Fig.~\ref{fig:pr_curve}.
\end{itemize}
The default threshold for test set was 0.98, so the final classifier data output would have feasible size to look through.

Table~\ref{tab:test_metrisc} shows that the Blender model never shows the best performance by any metrics.
However, we found that the Blender model shows better performance for our particular use-case.
Figure~\ref{fig:pr_curve} shows that the Blender model shows better dynamics for small recall / large precision region, which is what we are targeting.

\subsubsection{Early experiments and model selection}\label{sec:early-experiments}

After we trained the initial version of Random~Forest and CatBoost, we made an experiment of applying them to the target dataset.
To get an estimation of efficiency of the models on the target dataset, we evaluated them and, for each model, randomly selected a hundred objects classified as flares.
We visually classified these small datasets and made two observations: precision of the models are much smaller than what we see in Table~\ref{tab:test_metrisc}, and that a lot of mis-classification are from the asteroid occultation.
For the final estimate of precision for this experiment we drop all asteroid occultations, since they are easy to filter out on the post-filtering step (see Section~\ref{sec:asteriod-occultations}).
The precision for the CatBoost was found to be 0.030, for Random~Forest 0.054.
Those precision values would not allow us to efficiently classify the objects, so we decided to make an ensemble model.
We tried both mean-score value (precision 0.052) and minimum-score value (precision 0.077).
We chose to use the latest model, Blender, which was described by the previous section.

We re-run this experiment multiple times when changing input dataset (we initially used ZTF DR14) and collected a dataset of both flares and non-flares.
Since we also found that a lot of cases of mis-classification are caused by the various photometric pipeline problems, we used this small labeled dataset to built an ML classification model for the post-filtering step.

\begin{table}
  \centering
  \caption{Evaluation metrics on validation and test sets (classification threshold $0.98$).}
  \label{tab:test_metrisc}
  \setlength{\tabcolsep}{3pt}
  \footnotesize
  \begin{threeparttable}
  \begin{adjustbox}{max width=\linewidth}
  \begin{tabular}{@{}l
      S S
      S S
      S S@{}}
    \toprule
    & \multicolumn{2}{c}{Random Forest}
    & \multicolumn{2}{c}{CatBoost}
    & \multicolumn{2}{c}{Blender} \\
    \cmidrule(lr){2-3}\cmidrule(lr){4-5}\cmidrule(lr){6-7}
    Metric & {Validation} & {Test} & {Validation} & {Test} & {Validation} & {Test} \\
    \midrule
    Precision          & 0.997 & \bfseries 0.961 & 0.997 & 0.929             & 0.999 & 0.952 \\
    Recall             & 0.622 & 0.495           & 0.732 & \bfseries 0.525    & 0.613 & 0.404 \\
    F1 score           & 0.766 & 0.653           & 0.844 & \bfseries 0.671    & 0.759 & 0.567 \\
    Average precision  & 0.985 & \bfseries 0.925 & 0.993 & 0.919             & 0.991 & 0.924 \\
    \bottomrule
  \end{tabular}
  \end{adjustbox}
  \begin{tablenotes}[flushleft]\footnotesize
    \item Bold indicates the best \emph{test} value per metric.
  \end{tablenotes}
  \end{threeparttable}
\end{table}

\subsection{Post-filtering}\label{sec:post-filtering}

Early experiments showed that among the candidates the Blender classifier mis-classifies there are different astronomical phenomena, observational and pipeline artifacts: periodic variables, asteroid occulations,  defocusing in crowded fields, image artifacts, etc (see Section~\ref{sec:visual-inspection} for more details).
To improve the efficiency of the manual classification, we apply three post-filtering stages to the candidates selected by the light-curve classification model.

\subsubsection{Asteroid occultations}\label{sec:asteriod-occultations}
First, we check for nearby minor planets using the MPChecker service.
We set search radius to be 15 arcseconds and query for time of the maximum brightness.
The number of detected signals from known asteroids is 7582, which $\sim15\%$ of the total number of candidates given by the light-curve classifier.

Since this step required access to the MPChecker service, it may not be done in the beginning on the whole target dataset of dozen millions of objects.

\subsubsection{Classification model based on image quality metrics}

A lot of the candidates come from crowded Galactic fields, where photometry of stars can be contaminated by nearby objects if the image seeing changes between observations.
We have already detected such kind of bogus variability in previous researches, \cite{Malanchev_2021} gave an example of a seeing change causing stochastically-looking behavior of a light curve, and \cite{Voloshina_2024} demonstrated the contamination of flare-like light-curves with these kind of artifacts.
That made us believe that if brightening is correlated with seeing change then the detection is likely to be bogus \citep{2025RNAAS...9..156K}.
Since this correlation can be tricky to simulate, we built a machine-learning algorithm on a few observation metadata features, and trained the model on data labeled by \cite{Voloshina_2024} and Section~\ref{sec:early-experiments}.

We chose to use point spread function full width on the half maximum\footnote{We call this value ``seeing'' in this section to avoid a possible confusion with full the width on the half maximum of the light-curve shape we introduce in Section~\ref{sec:physical-parameter}.} and sharpness of the image source.
We used a linear regression binary classifier based on six features: seeing and sharpness at the first observation and the maximum light, and the difference between these two seeing and sharpness values.
All features were scaled to have zero mean and unity variance.

These exposure metadata are not available in the bulk-downloadable ZTF DRs, so we query the IRSA light-curve API\footnote{\url{https://irsa.ipac.caltech.edu/docs/program_interface/ztf_lightcurve_api.html}} for each specific object.
For training we used 607 non-flare light-curves and 120 flare light-curves.
2617 candidates for which the metadata was not available were dropped from the target sample.

Using a threshold of positive class probability equal to 0.9998, 4763 candidates were classifier as flare candidates and proceed to the next post-filtering step.

\subsubsection{Color-based M-dwarf selection}

During the visual analysis of the candidates we found that objects which are bluer than M~dwarfs in quiescence never show real flare activity.
Since we had more than 4~thousand candidates to go, we decided to make focus of this paper on M-dwarf flares only.
Using Pan-STARRS1~DR2 colors we kept all objects with $r-i$ color larger than 0.42.

\subsection{Visual inspection}\label{sec:visual-inspection}

\begin{figure*}
    \centering
    \includegraphics[width=0.8\textwidth]{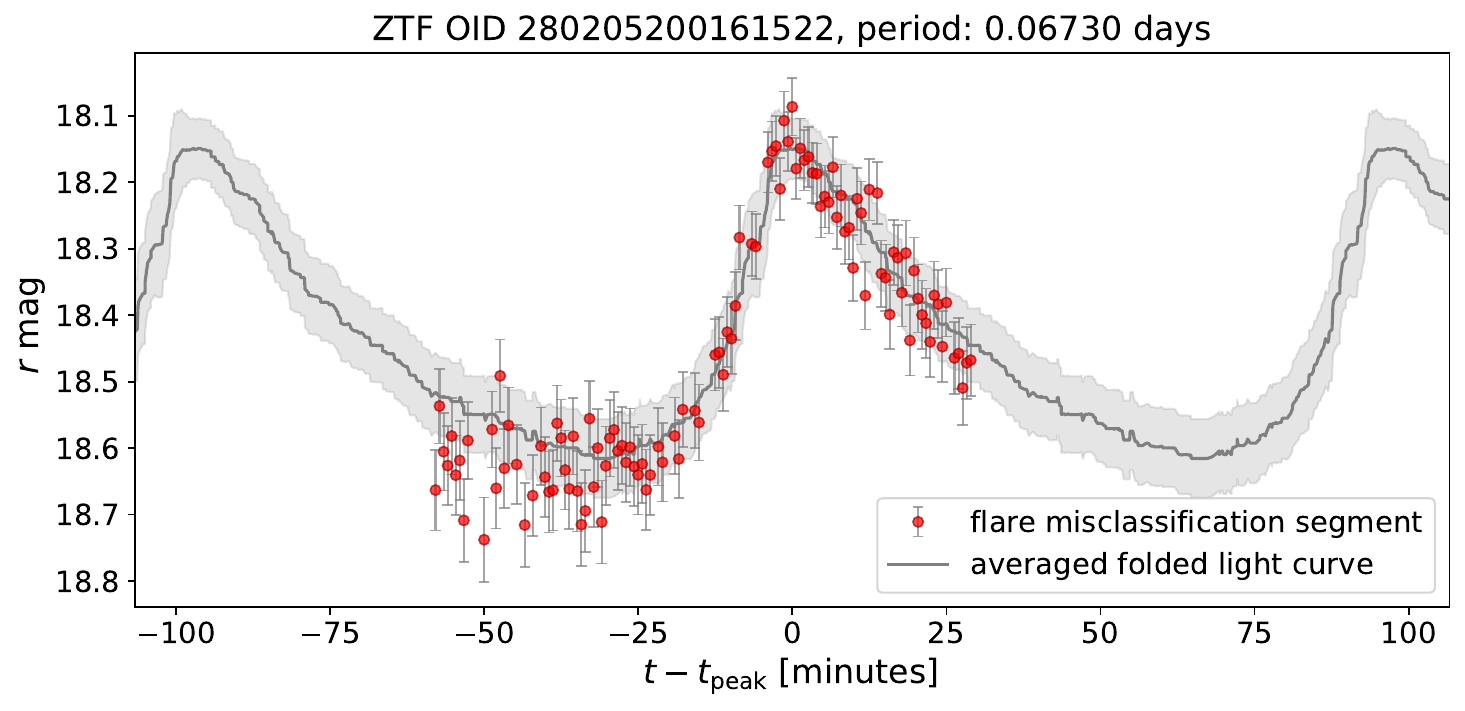}  
    \caption{Light curve of a $\delta$-Scuti star misclassified by the classification pipeline. The grey line shows the period-folded light curve of all observations, smoothed with a median filter. The light grey shaded area indicates the average deviation. Red points highlight high-cadence observations only, which were incorrectly classified as a stellar flare by our classification algorithm.}
    \label{fig:misclassification-dsct}
\end{figure*}

\begin{figure}
    \centering
    \begin{subfigure}[b]{0.48\textwidth}
        \includegraphics[width=\linewidth]{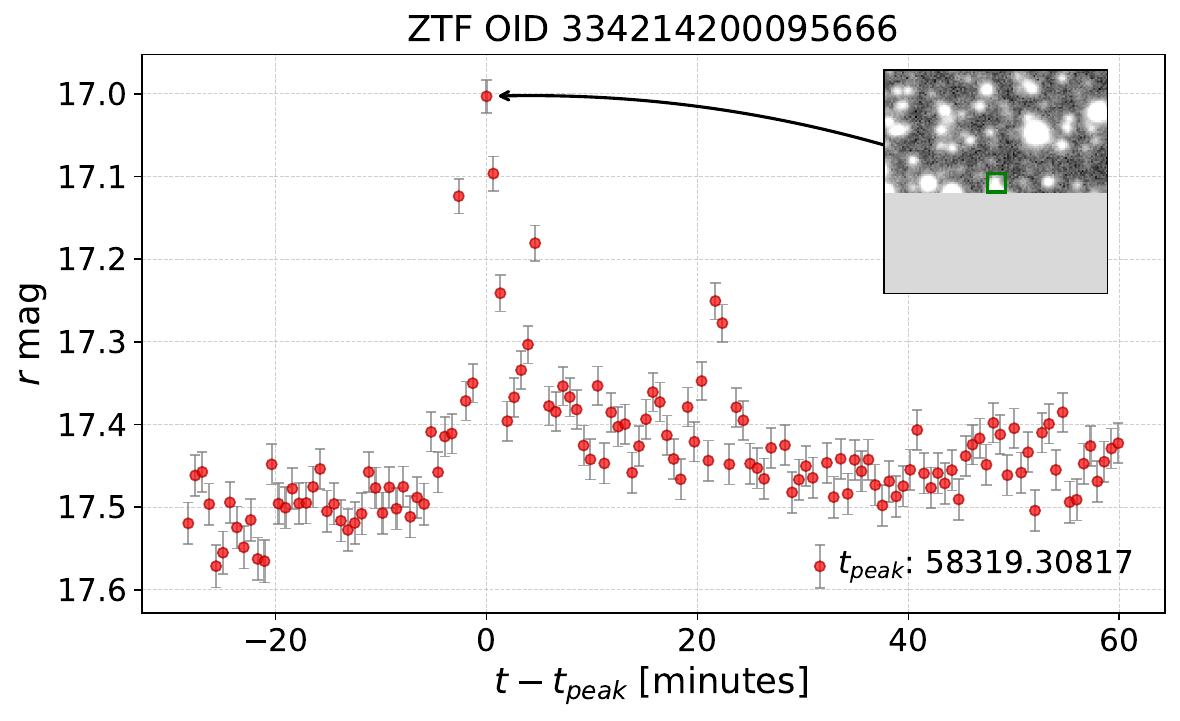}
        \caption{Defocusing}
    \label{fig:artefacts-defocusing}
    \end{subfigure}
    \hfill
    \begin{subfigure}[b]{0.48\textwidth}
        \includegraphics[width=\linewidth]{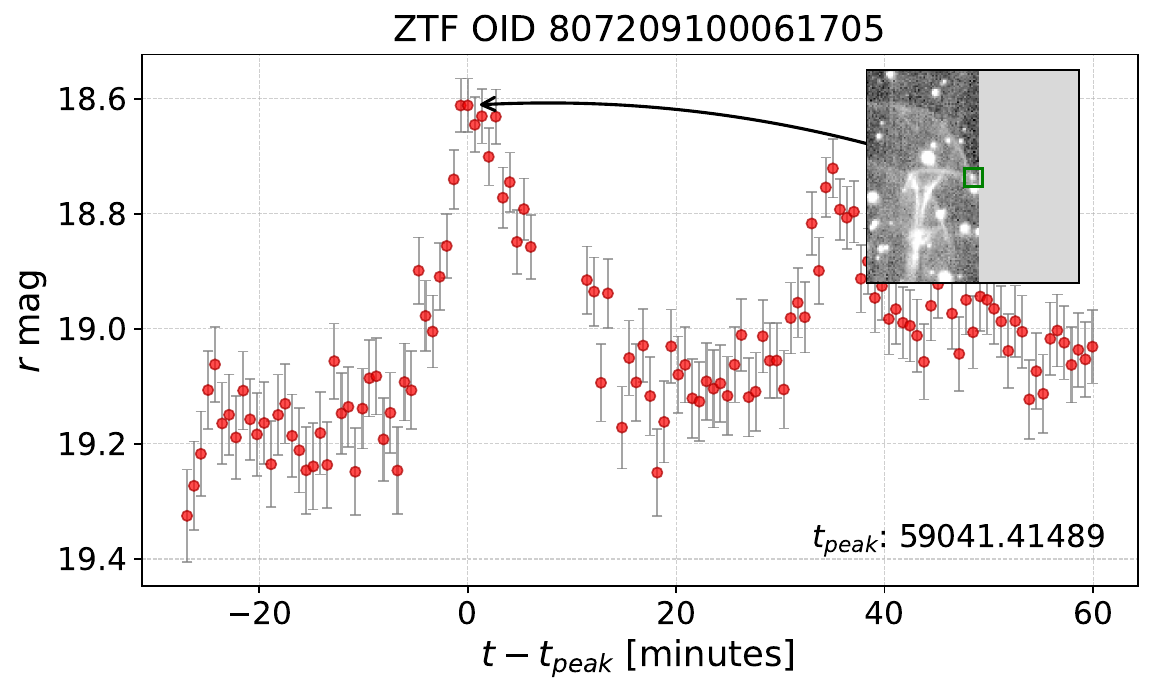}
        \caption{Ghost}
    \label{fig:artefacts-ghost}
    \end{subfigure}
    \hfill
    \begin{subfigure}[b]{0.48\textwidth}
        \includegraphics[width=\linewidth]{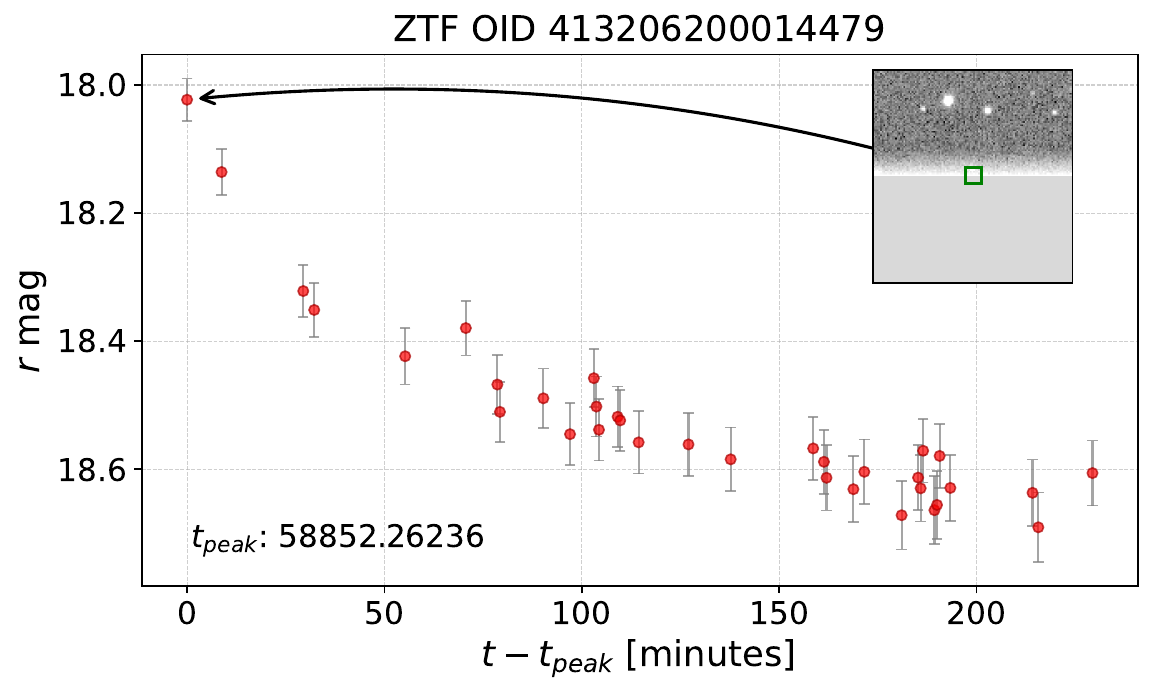}
        \caption{Frame edge}
    \label{fig:artefacts-frame-edge}
    \end{subfigure}
    \caption{Photometric artefacts that mimic stellar flares. The panels show short-duration, asymmetric brightenings -- fast rise followed by slower decay -- produced by instrumental/systematic effects: defocusing and blending of neighbouring PSFs (\ref{fig:artefacts-defocusing}); ghosts from internal reflections (\ref{fig:artefacts-ghost}); and proximity to the frame edge (\ref{fig:artefacts-frame-edge}). Such events are often misclassified by our classification pipeline as stellar flares; for a broader overview and publicly available ZTF DR3 dataset of such artefacts, see \citet{artefacts_vasha_2025_arxiv}.}
    \label{fig:artefacts}
\end{figure}

During the process of candidates analysis obtained after application of the classification model and the postfiltering we visually scrutinized light curves to sort out such misclassifications: 

\begin{itemize}
    \item \textbf{Periodic variable stars}. 
    Figure~\ref{fig:misclassification-dsct} presents a representative example of a misclassified light curve.
    Such misclassifications are primarily caused by limitations in the observing cadence in combination with the intrinsic periodicity of the source.
    Due to the sparse and irregular sampling of the light curves, only short segments of the periodic variability are observed. These segments can mimic the shape of a flare, particularly when they capture a rapid rise and decay phase.
    Consequently, in the absence of a full-phase light curve, it becomes difficult to distinguish periodic variables from genuine flaring events based solely on partial observational data.
    \item \textbf{Photometric artefacts}.
    An example of misclassified light curves caused by photometric artifacts is shown in Fig.~\ref{fig:artefacts}.
    Instrumental effects can produce short-lived apparent brightenings with the same asymmetric morphology -- fast rise followed by a slower decay -- typical of stellar flares. 
    For a broader overview and a publicly available dataset of such artefacts in ZTF DR3, see \citet{artefacts_vasha_2025_arxiv}.
    In our data the main contributors are:
    \begin{itemize}
        \item \textbf{Defocusing} (see Fig.~\ref{fig:artefacts-defocusing}):
        overlapping PSFs of nearby stars (blending) broaden and reshape the point-spread function. Small changes in seeing or focus can temporarily increase the contribution of neighbouring sources to the target aperture, producing an apparent impulsive brightening followed by a gradual return, thus mimicking a flare-like rise-decay profile.
        \item \textbf{Ghosts} (see Fig.~\ref{fig:artefacts-ghost}): ghosts, caused by internal reflections within lenses, filters, or their barrels, often appear as a diffuse smudge around the target; their motion into and out of the aperture creates transient brightenings with asymmetric morphology. 
        \item \textbf{Frame edge} (see Fig.~\ref{fig:artefacts-frame-edge}): frame edge effects occur when the target lies near the detector boundary, where small pointing shifts can move part of the PSF in or out of the frame or aperture, producing step-like or spike-like flux changes that can mimic flares.
    \end{itemize}
    \item \textbf{Symmetric light curves.}
    We also found a light curve with a nearly symmetric shape \textbf{ZTF OID 805205100020986}, whose morphology is more consistent with alternative phenomena e.g. self-lensing binaries \citep{Crossland_2024} than with a genuine flare.
\end{itemize} 

All described misclassifications arise because, in the absence of complementary information, the light-curve morphology alone is often insufficient to reliably distinguish genuine stellar flares from unrelated periodic variability, instrumental artefacts, or other transient phenomena.

\begin{figure*}
    \centering
    \includegraphics[width=\textwidth]{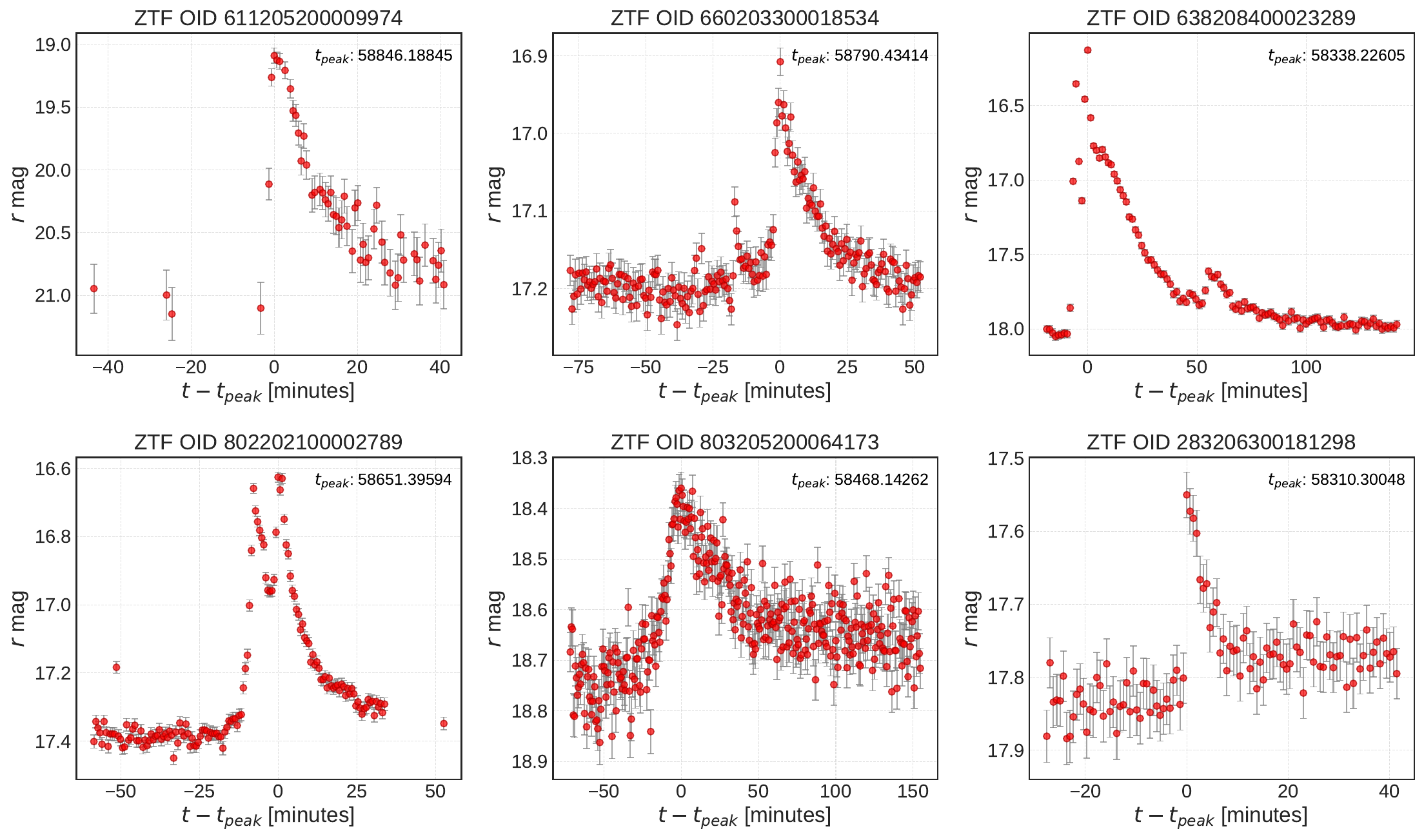}  
    \caption{Representative examples of found stellar flare activity in M dwarf stars. The light curves demonstrate the diverse morphology of flare events, ranging from simple single-peak flares to complex multi-peak structures, with amplitudes varying from subtle brightness increases to dramatic outbursts exceeding several magnitudes.}
    \label{fig:lightcurves}
\end{figure*}

\vspace{1em}
\hrule
\vspace{1em}

After applying the described pipeline, we obtained 2 898 candidates, and classified 1235 of them as M-dwarf flares. Six of these flares were observed in two ZTF fields and appeared as two different light curves in our dataset. This changes the total number of unique flares to {\textbf {1229}}. 

\section{Results}\label{sec:results}

In this section we analyze the population of flaring stars and physical properties of flares. We build an M-dwarf flare catalog (Section~\ref{sec:crossmatching}). We use this catalog to estimate a conditional distribution of flare frequency over spectral sub-classes (Section~\ref{sec:spec-class}) and vertical distance from the Galactic plane (Section~\ref{sec:spatial-distr}). Finally, we provide various physical parameters based on flare light curves, such as flare duration and its total energy (Section~\ref{sec:physical-parameter}).

\subsection{M-dwarf population in ZTF}\label{sec:crossmatching}

We selected a dataset of M-dwarf candidates from the entire ZTF DR17 data set in the r-band using Gaia and Pan-STARRS data using LSDB\footnote{\url{https://lsdb.io}} package~\citep{CaplarLSDB}.
First, we cross-matched all ZTF DR17 r-band objects ($\sim2.2$~billion) with Pan-STARRS1 DR2 object catalog, selecting the nearest neighbor within 1 arcsecond.
Following \citealt{Kowalski_2009} we filtered by $(r-i) > 0.42$ and $(i-z) > 0.23$ Pan-STARRS colors and magnitude errors in these bands to be smaller than 0.1.
We use this sample for color-based spectral classification in Section~\ref{sec:spec-class}.
We further refer to this catalog as \PSmatched{}, it contains 880,640,576 objects.

Next, we cross-matched \PSmatched{} with Gaia DR3 catalog, again selecting the nearest neighbor within 1 arcsecond.
We filter Gaia objects to have an upper limit of effective temperature (based on RP/BP spectra modelling) to be below 3800 kelvins, keeping M and later classes only~\citep{effective_temp}.
For the further analysis we also need access to distances, so we join the result catalog with Gaia Early DR3 distances catalog~\citep{Gaia_EDR3_dist_par}.
Since we rely on distances in further analysis we filter the catalog to have parallax to parallax error ratio greater than three.
This catalog is much smaller than the previous one and contains 27,420,797 objects.
We further refer to this catalog as \Gaiamatched{}.

\subsection{Spectral classification}\label{sec:spec-class}

The primary analysis is conducted using the effective temperature ($T_\mathrm{eff}$) from Gaia. In the final Table~\ref{tab:final-table}, we provide the spectral classification based on Gaia $T_\mathrm{eff}$ when available. If Gaia data is not available, we present the spectral subclass derived from PS1 photometry, without applying any dust correction.

\begin{table*}
\centering
\renewcommand{\arraystretch}{1.35}
\caption{Preview of the final table of flares parameters. The full table will be available online upon publication.}
\label{tab:final-table}
\label{tab:sample_flares}
\begin{tabular}{lrrrrlrrrrl}
\toprule
\textbf{ZTF DR OID} & 
\thead{$\bm{\alpha}$, \\ deg} & 
\thead{$\bm{\delta}$, \\ deg} & 
\thead{\textbf{distance}, \\ pc} & 
\thead{$\bm{A_r}$$^{1}$, \\ (mag)} & 
\thead{$\bm{t}_\mathrm{\bf peak}$$^{2}$, \\ MJD$-$58000} & 
\thead{\textbf{FWHM}$^{2}$, \\ hours} & 
\thead{\textbf{amplitude}$^{2}$, \\ $\Delta$ mag} & 
\textbf{n points}$^{2}$ & 
\thead{\textbf{spectral} \\ \textbf{class}} \\
\midrule
736211200022871 & 0.5619 & 49.3449 & $361.96^{+25.37}_{-28.37}$ & $0.1570_{-0.0262}$ & 2042.4119 & 6.57 & 0.632 & 56 & M2 \\
806208200015545 & 0.2700 & 61.6103 & $132.18^{+0.77}_{-0.97}$ & 0.0000 & 1473.1295 & 2.02 & 0.621 & 54 & M3 \\
806212300069862 & 0.5732 & 62.2869 & $^{\dagger}1949.39^{+1871.13}_{-774.08}$ & $0.9577^{+0.6199}_{-0.2250}$ & 2032.4054 & 0.93 & 1.626 & 26 & M4 \\
806216100027505 & 1.0642 & 65.6713 & $645.17^{+102.19}_{-82.13}$ & $1.4394^{+0.5219}_{-0.0518}$ & 2032.4363 & 7.99 & 0.525 & 41 & M3 \\
833208300019023 & 0.9214 & 67.6663 & $^{\dagger}1407.72^{+492.90}_{-305.74}$ & $3.5330^{+0.0262}_{-0.1047}$ & 2038.4438 & 7.34 & 0.503 & 30 & M4 \\
\addlinespace
\multicolumn{10}{c}{\ldots} \\
\bottomrule
\end{tabular}

\vspace{3pt}
\parbox{0.95\textwidth}{{\footnotesize
$^{1}$~For sources with reliable geometric distances, the three-dimensional Milky Way dust map \texttt{Bayestar19} \citep{Green_2019} was used to estimate extinction. 
If such distance information was unavailable, extinction values were taken from the Galactic Dust Reddening and Extinction map by \citealt{2011ApJ...737..103S}.\\
$^{2}$~The peak time, FWHM, amplitude, and number of points were obtained from parametric flare-profile fits. 
For sparsely sampled light curves, the peak time corresponds to the epoch of minimum observed magnitude, 
and the amplitude was computed as the difference between this minimum and the quiescent magnitude inferred from the model. 
FWHM values were measured only for events with sufficient data to construct a reliable flare profile.\\
$^{3}$~Spectral classification is based on \textit{Gaia}~DR3 effective temperature ($T_\mathrm{eff}$) when available. 
If no \textit{Gaia} information was present, the spectral subclass was derived from \textit{PS1} photometry without applying any dust correction.\\
$^{\dagger}$~Objects with $\mathrm{Plx}/\mathrm{e_{Plx}} < 5$ according to \textit{Gaia}~DR3 parallax estimates.
}}
\end{table*}

We used two methods to evaluate a spectral subclass for two different subsets of flares.
The first subset consists of the stars from \Gaiamatched{} with available effective temperature, thus we are using ratios from \cite{Malkov_2020} to define a star's spectral subclass.
The total amount of flaring stars in this sample -- 680 (we drop recurrent flares from the same source and take into account only total number of flaring stars).
The second one consists of the stars from \PSmatched{} sample with available photometric data (the total amount of flaring stars -- 1212).

To estimate the spectral subclass of each star, we use the probabilistic classification approach introduced by \citet{Kowalski_2009}, which models the distribution of M-dwarfs in the ($r-i$, $i-z$) colour–colour diagram as a set of two-dimensional Gaussians. Each spectral subtype $\mathcal{M}$ (from M0 to M7) is characterized by a mean vector $\boldsymbol{\mu}_\mathcal{M}$ and a covariance matrix $\Sigma_\mathcal{M}$.

For a given star with colour vector $\boldsymbol{x} = (r-i,\, i-z)$, the likelihood of belonging to subclass $\mathcal{M}$ is computed as:
\begin{equation}
 p_\mathcal{M}(\boldsymbol{x}) = \frac{1}{2\pi \sqrt{|\Sigma_\mathcal{M}|}} \exp\left( -\frac{1}{2} (\boldsymbol{x} - \boldsymbol{\mu}_\mathcal{M})^T \Sigma_\mathcal{M}^{-1} (\boldsymbol{x} - \boldsymbol{\mu}_\mathcal{M}) \right)
\end{equation}

The spectral subclass is then assigned based on the maximum probability $p_\mathcal{M}$ across all $\mathcal{M}$.

\begin{figure}
\includegraphics[width=\linewidth]{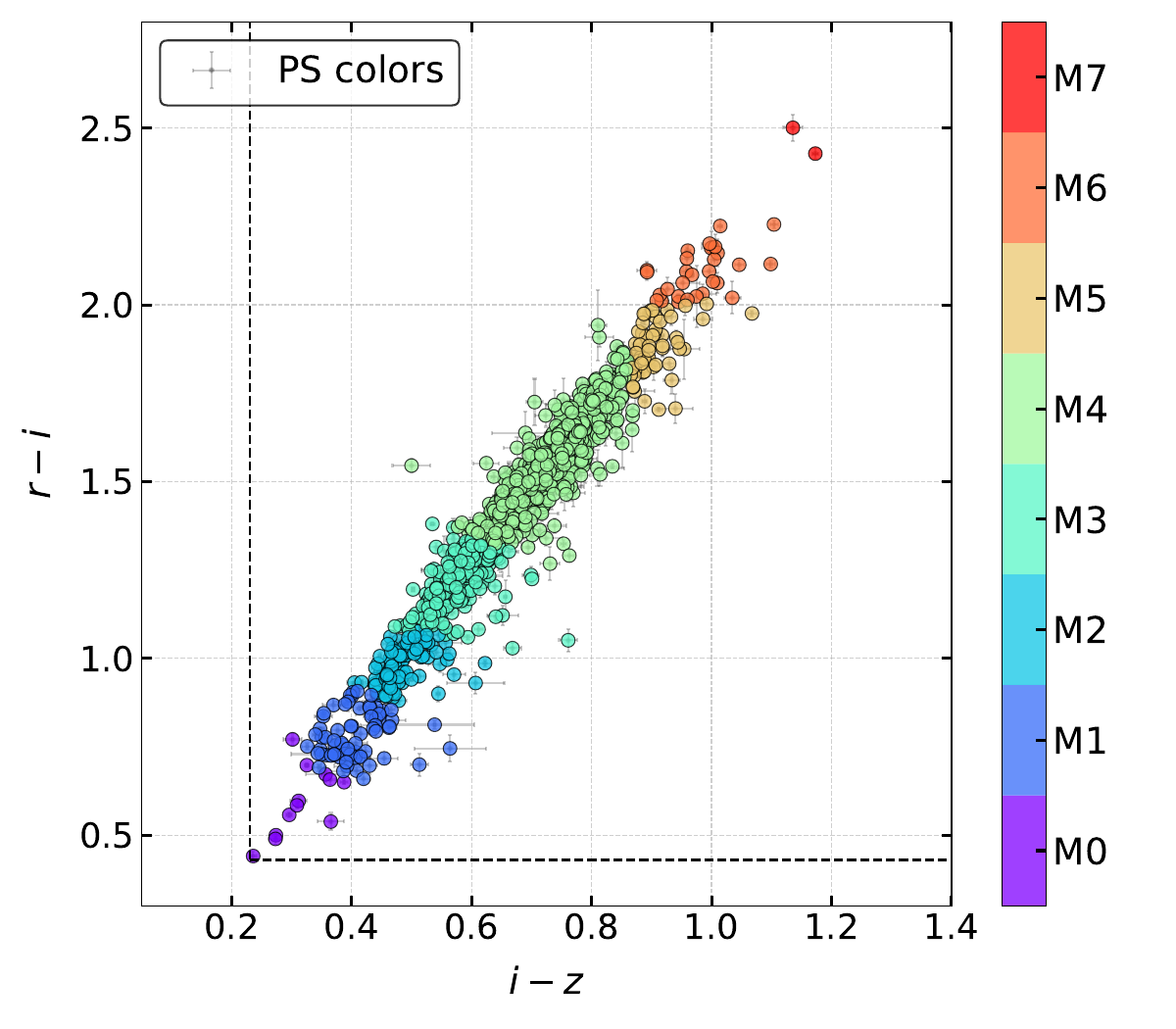}
\caption{Colour–colour diagram for \PSmatched{} sample.
The distribution of stars in the (r-i, i-z) plane is shown, with colored regions indicating the loci corresponding to spectral subclasses M0-M7 used for photometric classification in this work.}
\label{fig:flares-spec-class}
\end{figure}

\begin{figure*}
  \centering
  \begin{subfigure}{.5\textwidth}
    \centering
    \includegraphics[width=.8\linewidth]{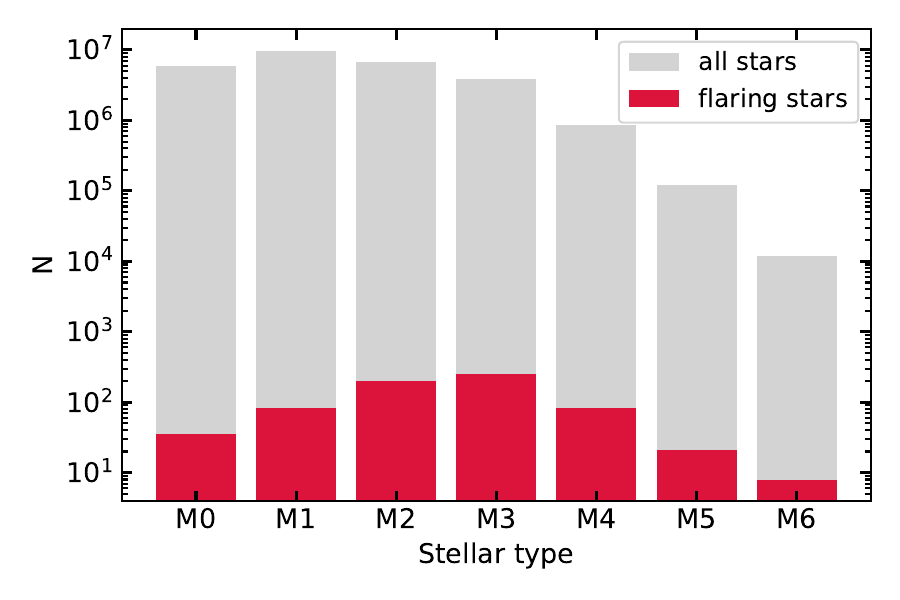}
    \caption{Number of flares per spectral subclass.}
    \label{fig:sfigg1}
  \end{subfigure}%
  \begin{subfigure}{.5\textwidth}
    \centering
    \includegraphics[width=.8\linewidth]{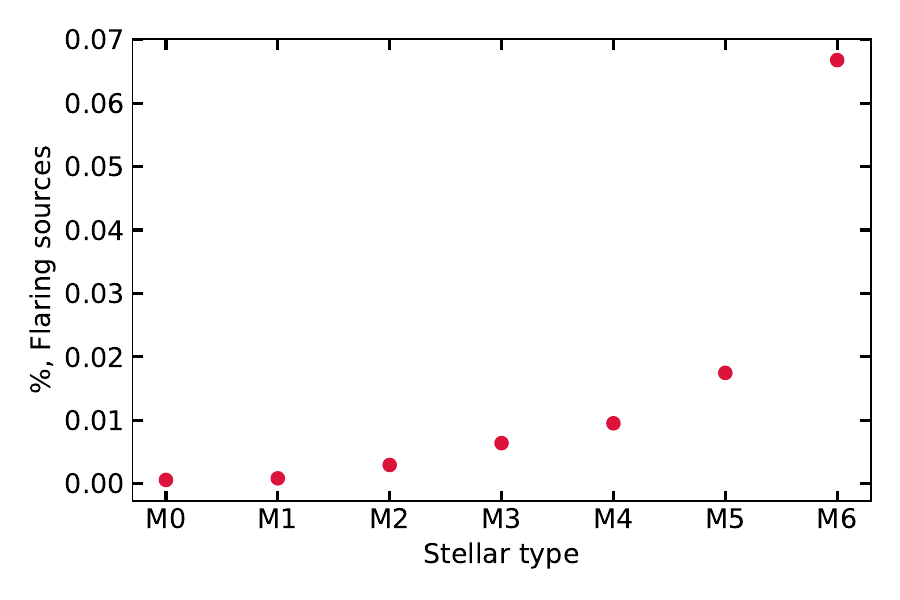}
    \caption{Fraction of flaring sources per subclass.}
    \label{fig:sfigg2}
  \end{subfigure}
  \caption{
    Statistics of flaring sources by spectral type for the \Gaiamatched{} sample.
    (\ref{fig:sfigg1}) Total number of stars of each M-dwarf subclass (grey) and those exhibiting flares (red).
    (\ref{fig:sfigg2}) Fraction of flaring sources relative to the total number of stars per spectral subclass.
  }
  \label{fig:fig1}
\end{figure*}

\begin{figure*}
\begin{subfigure}{.5\textwidth}
  \centering
  \includegraphics[width=.8\linewidth]{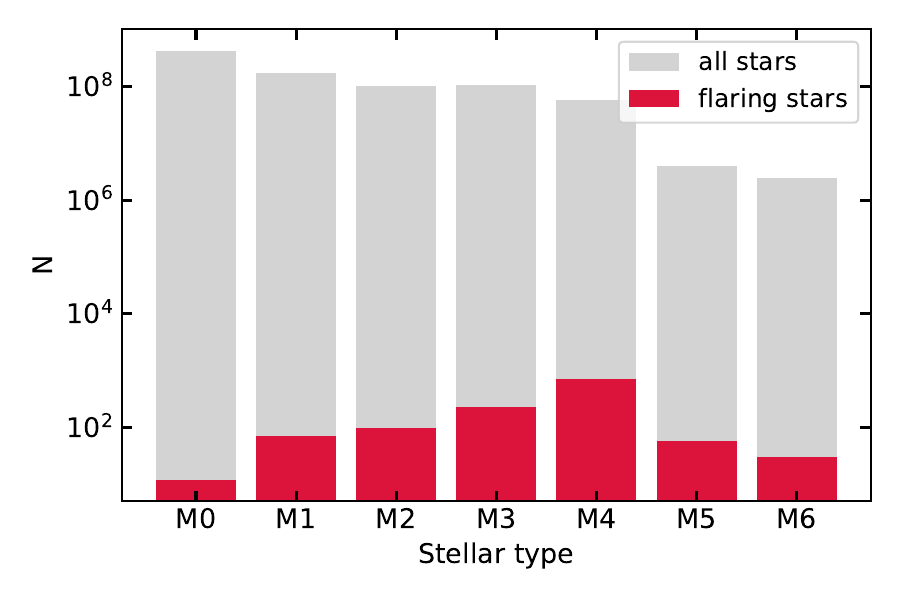}
  \caption{}
  \label{fig:sfig1}
\end{subfigure}%
\begin{subfigure}{.5\textwidth}
  \centering
  \includegraphics[width=.8\linewidth]{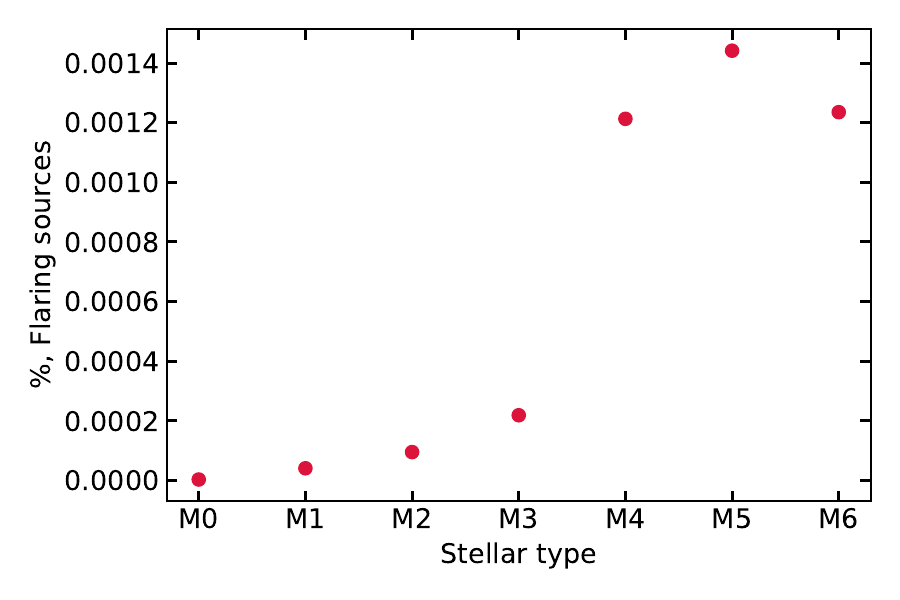}
  \caption{}
  \label{fig:sfig2}
\end{subfigure}
\caption{Statistics of flaring sources by spectral type for the \PSmatched{} sample. (\ref{fig:sfig1}) Total number of stars of each M-dwarf subclass (grey) and the number of stars exhibiting flares (red).
(\ref{fig:sfig2}) Fraction of flaring sources relative to the total number of stars of the corresponding subclass.}
\label{fig:fig2}
\end{figure*}

The result of such classification is shown in Fig.~\ref{fig:flares-spec-class}.
We present comparison between the number of stars of the certain spectral subclass versus corresponding number of flares for both subsets (see Fig.~\ref{fig:sfigg1} and Fig.~\ref{fig:sfig1}) as well as a comparison of flares frequencies per spectral subclass (Fig.~\ref{fig:sfigg2} and Fig.~\ref{fig:sfig2}).
Based on the analysis of \Gaiamatched{} there is a clear increasing trend between a flare frequency and a spectral subtype with a steep rise near M4-M5 where stars, tentatively, become fully convective and therefore generate more flares which is consistent with previous researches~\citep{tess_dr1_flares, DECam_flares}.
Oppositely there is no such evidence from \PSmatched{} sample analysis apparently due to the absence of color correction for the galactic reddening.
The distribution of spectral sub-classes is shifted towards the older spectral sub-classes, due to interstellar extinction.

\subsection{Physical parameter estimation}\label{sec:physical-parameter}

\subsubsection{Flare energy}\label{sec:flare-energy}
To conduct the energy estimation, we use the \Gaiamatched{} sample with available distance estimations. 
For the recurrent events, we estimate energy for each flare separately.
We exclude flares with unusual shapes or poorly sampled light curves (e.g., flares showing only a decay phase) and apply the quality cut on the parallax-to-error ratio:
$$\frac{\varpi}{\sigma_\varpi} > 3.$$

The final sample used for energy calculation contains \textcolor{blue}{655} flares. The energy estimation method follows the approach developed in~\cite{Voloshina_2024} and is summarized here for completeness. We assume that flare emission can be approximated by black-body radiation with $T_\text{flare} = 9000$\,K \citep{1992ApJS...78..565H}, yielding the bolometric luminosity:
\begin{equation}
    L_{\text{flare}}(t) = \sigma_\text{SB} T^4_\text{flare} \, d^2 \, \frac{F_r(t)}{B_r},
\end{equation}
where $F_r(t)$ is the extinction-corrected flux in the ZTF $r$-band, and $B_r$ is the band-averaged black-body intensity. Following \citet{Shibayama_2013} and \citet{Yang2017}, we assume the geometric correction factor is unity.

Observed magnitudes are converted to fluxes using 3D extinction maps~\citep{Green_2019}, and the light curves are fit with a modified semi-empirical flare model from \citet{Mendoza}, as described in~\cite{Voloshina_2024}. The total flare energy is then obtained by integrating the luminosity:
\begin{equation}
    E_{\text{flare}} = \int L_{\text{flare}}(t)\,\text{d}t.
\end{equation}

A power-law relation is found between flare duration (see detatils about calulation of flare duration $t_{\text{FWHM}}$ in Section~\ref{sec:flare-fit-params}) and bolometric energy estimation:
\begin{equation}
    E_{\text{flare}} = 10^{32.045} \times t_{\text{FWHM}}^{1.306},
\end{equation}
where $t_{\text{FWHM}}$ is the full width at half maximum (in minutes) derived from the best-fit model light curve.

\begin{figure}
\center{
\includegraphics[width=0.45\textwidth]{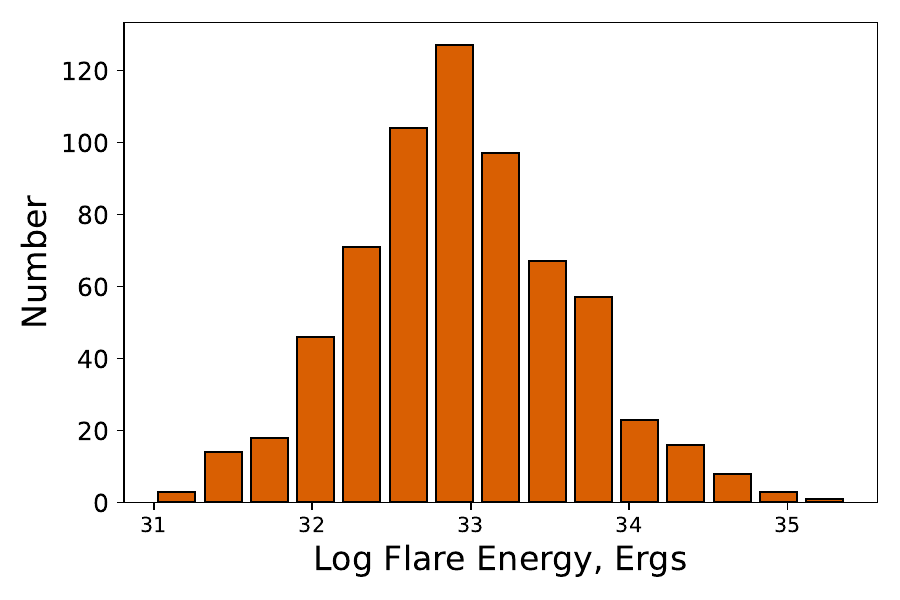}
}
\caption{Histogram showing the number of flare events per logarithmic energy bin.
The sample spans approximately $10^{31}-10^{35}$
erg, derived for 655 \Gaiamatched{}
flares with reliable distance estimates.}
\label{fig:energy-hist}
\vspace{0.2cm}
\end{figure}

\begin{figure}
\center{
\includegraphics[width=0.45\textwidth]{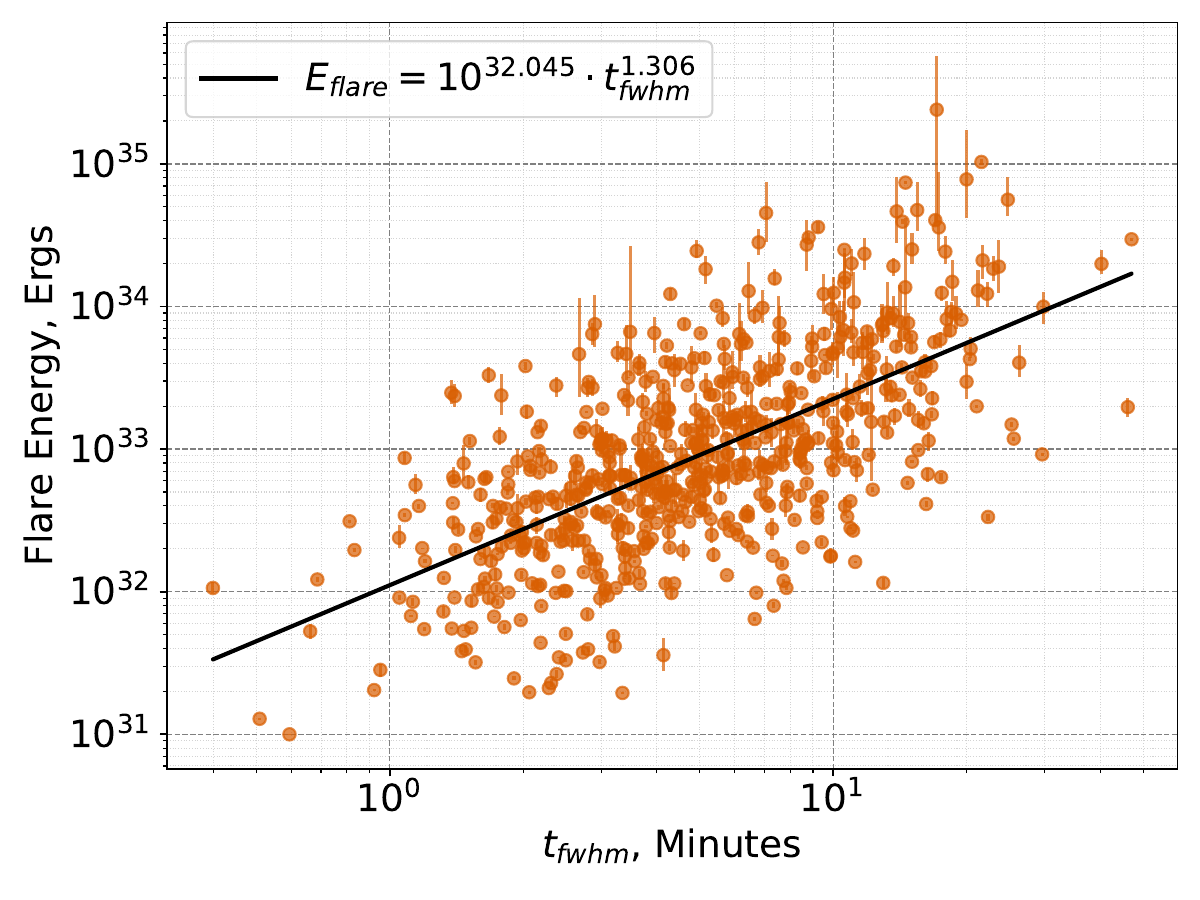}
}
\caption{Bolometric flare energy as a function of the full width at half maximum ($t_{\mathrm{fwhm}}$) of the modeled light curves. The best-fit power-law relation $E_{\mathrm{flare}} = 10^{32.045} t_{\mathrm{FWHM}}^{1.306}$.}
\label{fig:fwhm-scatter}
\vspace{0.2cm}
\end{figure}

\begin{table}
\centering
\renewcommand{\arraystretch}{1.2}
\begin{tabular}{ccc}
\toprule
\midrule
\textbf{OID} & \textbf{FWHM}, minutes & \textbf{E}, $10^{33}$ erg \\
\midrule
257207200014674 & 2.82 & $0.58 \pm^{0.07}_{0.06}$ \\[2px]
257207200014907 & 3.13 & $0.88 \pm^{0.03}_{0.03}$ \\[2px]
259205400044242 & 7.53 & $4.25 \pm^{1.75}_{1.01}$ \\[2px]
259207300021409 & 2.99 & $1.16 \pm^{0.05}_{0.04}$ \\[2px]
280211400109529 & 2.18 & $0.68 \pm^{0.02}_{0.03}$ \\[2px]
\multicolumn{3}{c}{$\cdots$} \\[2px]
\bottomrule
\bottomrule
\end{tabular}
\caption{
Preview of bolometric energy estimations for a sample of flare candidates. The full table will be available online upon publication. Energy errors reflect uncertainties in geometric distance estimates (see Sec.~\ref{sec:flare-energy}).
}
\label{tab:energy-preview}
\end{table}

\subsubsection{Flare fit parameters}\label{sec:flare-fit-params}
For each flare profile, we evaluated a set of parameters based on the semi-empirical model from~\cite{Mendoza}, including the amplitude, full width at half maximum ($t_{\text{FWHM}}$), and the number of observations during the flare.

In cases where a light curve contains multiple flares, we treated each flare separately by isolating and analyzing them independently.
We performed a parametric fit to each flare profile to obtain a continuous representation of the flare.
The amplitude was calculated as the difference between the maximum and minimum of the modeled flux.
To estimate the FWHM, we measured the time interval between points where the modeled flux equals half of the amplitude.

To determine the number of data points associated with a flare, we used the following criterion: any point in the light curve with observed flux exceeding the quiescent stellar flux ($f^*$, as defined by the model) by more than $3\sigma$ -- where $\sigma$ is the average observational uncertainty over the selected segment -- was classified as part of the flare.

The resulting parameters for the representative subset of flares are listed in Table~\ref{tab:final-table}.

\subsection{Spatial distribution}\label{sec:spatial-distr}
Magnetic activity in low-mass stars provides a valuable tracer of stellar age and Galactic structure. 
Because vertical distance from the Galactic plane $|z|$ correlates with stellar age through kinematic heating, the fraction of magnetically active or flaring M dwarfs can serve as a probe of the temporal evolution of magnetic activity across the Galactic disk. 
Our goal is therefore to quantify how the proportion of flaring stars -- rather than the rate of individual flare events -- varies with Galactic height.

For each observing field and vertical bin, we measured the total number of stars, $N_{\mathrm{star}}$, and the number of stars exhibiting at least one detected flare, $N_{\mathrm{flare}}$. The data are naturally described by a binomial process:
\begin{equation}
N_{\mathrm{flare},\, i,\,f} \sim \mathrm{Binomial}\left(N_{\mathrm{star},\,i,\,f},\,\, p_{i,\,f}\right),
\end{equation}
where $p_{if}$ is the probability that a randomly selected star in bin $i$ and field $f$ is flare-active. We emphasize that $p_{if}$ represents the likelihood that a star belongs to the flaring (magnetically active) population, not the instantaneous probability of an individual flare event.

We modeled the dependence of this probability on Galactic height using a binomial generalized linear model (GLM) with a logit link function. The logit function,
\begin{equation}
\mathrm{logit}(p_{i,\,f}) = \log \left( \frac{p_{i,\,f}}{1 - p_{i,\,f}} \right),
\end{equation}
maps probabilities in the range $0 < p_{if} < 1$ to the full real axis, allowing us to express the expected log-odds of flaring activity as a linear combination of predictors:
\begin{equation}
\mathrm{logit}\,p_{i,\,f} = \alpha + \beta \, \log |z_{i,\,f}| + \gamma_f.
\end{equation}

\begin{figure}
    \centering
    \includegraphics[width=0.48\textwidth]{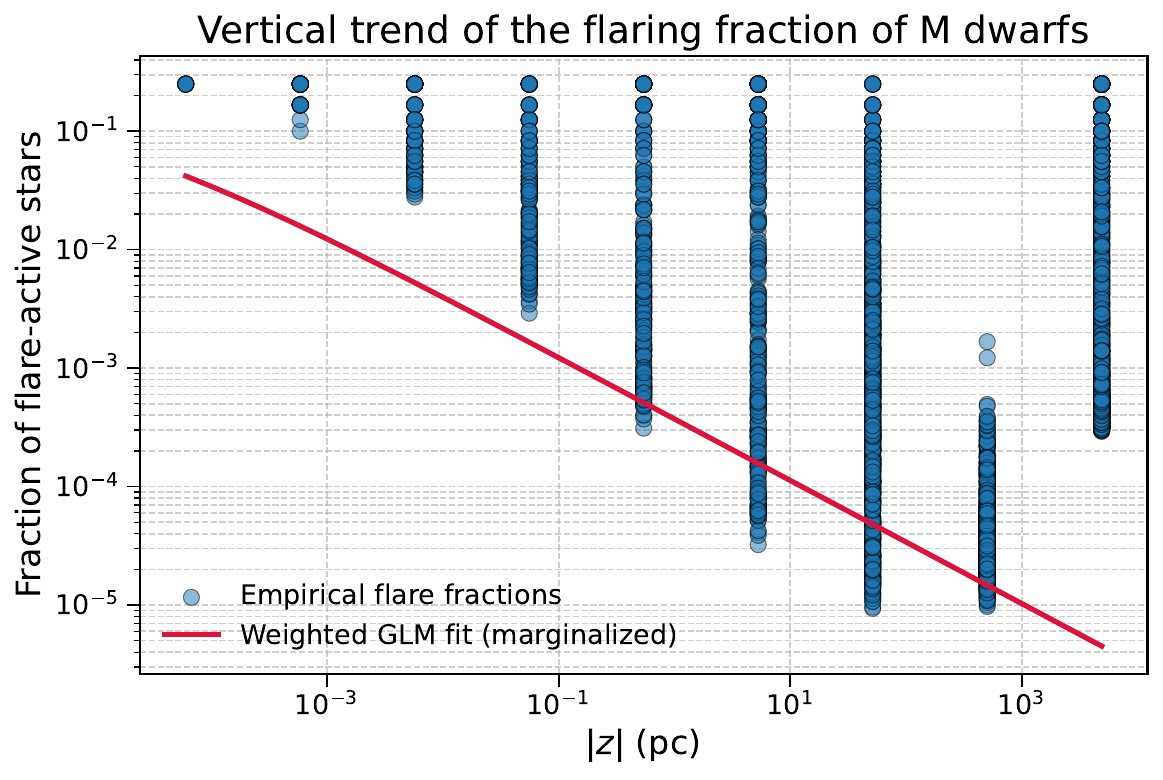}
    \caption{
        Vertical trend of the flaring fraction of M dwarfs.
        Blue points show empirical flare fractions per field,
        while the solid red curve represents the weighted binomial GLM prediction
        marginalized over fields and scaled by stellar counts.
        Both axes are logarithmic.
    }
    \label{fig:flaring_fraction}
\end{figure}

Fitting this model to the observed data yields a highly significant negative slope of
\begin{equation}
\beta = -0.52 \pm 0.03 \quad (p < 10^{-15}),
\end{equation}
indicating that the odds of a star being flare-active decrease by roughly a factor of three for every for each order-of-magnitude increase in $|z|$. 
This trend remains consistent when tested with Poisson and negative binomial GLMs, confirming that it is robust to the assumed statistical family. 
The observed decline of $p(|z|)$ with Galactic height reflects the progressive aging of stellar populations away from the midplane and the corresponding decay of magnetic activity with time.
Figure~\ref{fig:flaring_fraction} shows the vertical distribution of flare-active M dwarfs and the fitted generalized linear model (GLM) for the flaring fraction as a function of Galactic height. The blue points represent empirical flare fractions measured in individual fields, computed as the ratio $N_{\mathrm{flare}}/N_{\mathrm{star}}$ in each $|z|$ bin. 
The solid red curve shows the weighted binomial GLM prediction marginalized over fields and scaled by the total number of stars per field. 

\section{Conclusions}

In this work, we have constructed and presented the largest catalogue to date of M-dwarf flares from a ground-based survey, consisting of 1,229 unique flaring events identified in data from the Zwicky Transient Facility.
This catalogue provides a substantial, well-characterized sample for studying the energetic phenomena on low-mass stars.

Our results were achieved through a multi-stage pipeline trained on simulated examples to efficiently scan through billions of light curves and isolate flare events. 
Training examples were simulated by injecting flare templates from the TESS space telescope into actual ZTF quiescent light curves, ensuring the model learned to identify flares within the specific noise properties and cadence of the ZTF survey. 
To maximize the purity of the final catalogue, we used a post-filtering process that successfully removed a significant number of false positives by: (1) cross-match against minor planet ephemerides; (2) deploying a second ML model to filter out instrumental artifacts based on image quality metadata; and (3) using Pan-STARRS colors to select for M-dwarf candidates.
The developed pipeline is adaptable for detecting not only flares but also various other transient phenomena, such as short transients (for example, Fast Radio Bursts).

The resulting catalogue offers detailed characterizations of both the flares and their host stars. 
By leveraging photometric data from Pan-STARRS and astrometric data from Gaia, we analyzed the distribution of flare hosts across spectral subtypes.
Our analysis confirms previous findings, revealing a clear increase in flare frequency with later spectral types, with a significant peak around the M4–M5 subtypes. 
This observation is consistent with the theoretical transition to fully convective interiors in M-dwarfs, which is thought to enhance dynamo action and magnetic activity.
Furthermore, for a high-quality subset of 655 flares with reliable Gaia distance measurements, we estimated bolometric energies and durations. 

Establishing robust methodologies for flare detection and characterization is crucial for the era of upcoming large-scale transient surveys. 
Our developed pipeline and post-filtering strategy provide a blueprint for efficiently processing the unprecedented volumes of data expected from these next-generation facilities. Missions like the Vera C. Rubin Observatory's Legacy Survey of Space and Time (LSST), the Nancy Grace Roman Space Telescope, \textit{ULTRASAT}, and the Argus Array will each contribute unique capabilities to our understanding of the dynamic sky.
By demonstrating the successful extraction of a large and pure sample of M-dwarf flares from ZTF data, our research paves the way for even more comprehensive studies of stellar activity across the vast parameter space that these future surveys will unlock, ultimately enhancing our understanding of stellar evolution, exoplanet habitability, and the transient universe.

\section*{Data Availability}

The ZTF light-curve data underlying this article are available from the NASA/IPAC Infrared Science Archive\footnote{\url{https://irsa.ipac.caltech.edu/}}. The final catalogue of selected flares and their derived parameters will be made publicly available via GitHub\footnote{\url{https://github.com/snad-space/flare-classifier}} and the VizieR database upon publication.

\section*{Acknowledgements}
A.~Lavrukhina and M.~Pruzhinskaya acknowledge support from a Russian Science Foundation grant 24-22-00233, https://rscf.ru/en/project/24-22-00233/ for the analysis of M-dwarf population and their physical
properties.
Support was provided by Schmidt Sciences, LLC. for K.~Malanchev.
The computational resources were provided by Yandex Cloud\footnote{\url{https://yandex.cloud}} as part of a grant for scientific research. K.~Malanchev and E.E.O. Ishida acknowledge support from 2024 CNRS International Emerging Actions A100011.

\bibliographystyle{mnras}
\bibliography{example}


\appendix

\section{List of experts features for training}\label{sec:features}

We define light curve as a list of triples of magnitude $m_i$, its error $\delta_i$ and observation time $t_i$, where $i = 0..N-1$ and $N$ is the number of observations.

\paragraph{Mag min}

\paragraph{Bazin fit}
Five fit parameters and goodness of fit (reduced $\chi ^{2}$ of the Bazin function developed for core-collapsed supernovae:
$\displaystyle f(t)=A\frac{\mathrm{e}^{-(t-t_{0} )/\tau _{fall}}}{1+\mathrm{e}^{-(t-t_{0} )/\tau _{rise}}} +B,$ where $f(t)-$ flux observation

\paragraph{Median Buffer Range Percentage}
Fraction of points within $\displaystyle \mathrm{Median} (m)\pm q\times (\max (m)-\min (m))/2$
We used $n = 0.01$ in our analysis.

\paragraph{Amplitude}
The half amplitude of the light curve:
\begin{equation}
    \frac{\max(m) - \min(m)}{2}.
\end{equation}

\paragraph{Percent Difference Magnitude Percentile}
Ratio of $p$-th inter-percentile range to the median:
$\displaystyle \frac{Q( 1-p) -Q( p)}{\text{Median}( m)}$
We used $p = 0.05$ in our analysis.

\paragraph{$\eta^e$}
\cite{Kim_etal2014}
Generalisation of $\eta$for unevenly time series.
\begin{equation}\label{eq:eta_e}
    \eta^e \equiv (t_{N-1} - t_0)^2 \frac{\sum_{i=0}^{N-2} \left(\frac{m_{i+1} - m_i}{t_{i+1} - t_i}\right)^2}{(N - 1)\,\sigma_m^2}.
\end{equation}
Our version of this feature differs from \cite{Kim_etal2014}.

\paragraph{Anderson Darling Normal}
Unbiased Anderson–Darling normality test statistic:
$\displaystyle \left( 1+\frac{4}{N} -\frac{25}{N^{2}}\right) \times$
$\times \left( -N-\frac{1}{N}\sum\limits_{i=0}^{N-1} (2i+1)\ln \Phi _{i} +(2(N-i)-1)\ln (1-\Phi _{i} )\right) ,$ where $\Phi _{i\ } \equiv \Phi (( m_{i} \ -\ \langle m\rangle ) /\sigma _{m})$ is the commutative distribution function of the standard normal distribution, $N-$ the number of observations, $\langle m\rangle -$ mean magnitude and $\sigma _{m} =\sqrt{\sum\limits_{i=0}^{N-1}( m_{i} -\langle m\rangle )^{2} /( N-1) \ }$ is the magnitude standard deviation.

\paragraph{Standard Deviation}
Standard deviation of magnitude:
$\displaystyle \sigma _{m} \equiv \sqrt{\sum _{i} (m_{i} -\langle m\rangle )^{2} /(N-1)}$

\paragraph{Median}
We define $\median(x) \equiv Q_x(0.5)$.

\paragraph{Mean}
We define mean of the sample $\{x_i\}$ as
\begin{equation}
    \xmean \equiv \frac1{N} \sum_i x_i.
\end{equation}

\paragraph{Von Neummann $\eta$}
\cite{Kim_etal2014}
\begin{equation}\label{eq:eta}
    \eta \equiv \frac1{(N - 1)\,\sigma_m^2} \sum_{i=0}^{N-2}(m_{i+1} - m_i)^2.
\end{equation}

\paragraph{Median Absolute Deviation}
\cite{DIsanto_etal2016}
Median of the absolute value of the difference between magnitudes and their median:
\begin{equation}
    \median\left(|m_i - \median(m)|\right).
\end{equation}

\paragraph{Beyond $n$ Std}
\cite{DIsanto_etal2016}
The fraction of observations beyond $\mean \pm n\,\sigma_m$:
\begin{equation}
    \frac{\sum_i I_{|m - \mean| > n\sigma_m}(m_i)}{N}.
\end{equation}
We use $n = 1$ and $n = 2$ in our analysis.

\paragraph{Skew}
The skewness of magnitude distribution:
\begin{equation}
    G_1 \equiv \frac{N}{(N - 1)(N - 2)} \frac{\sum_i(m_i - \mean)^3}{\sigma_m^3}.
\end{equation}

\paragraph{Weighted Mean}
Weighted mean magnitude:
$\displaystyle \overline{m} \equiv \frac{\sum _{i} m_{i} /\delta _{i}^{2}}{\sum _{i} 1/\delta _{i}^{2}}$

\paragraph{Excess Variance}
Measure of the variability amplitude:
$\displaystyle \frac{\sigma _{m}^{2} -\langle \delta ^{2} \rangle }{\langle m\rangle ^{2}},$ where $\langle \delta ^{2} \rangle -$ mean squared error.

\paragraph{Stetson $K$}
\cite{Stetson1996}
Stetson K coefficient described light curve shape:
\begin{equation}
    K \equiv \frac{\sum_i\left|\frac{m_i - \bar{m}}{\delta_i}\right|}{\sqrt{N \sum_i\left(\frac{m_i - \bar{m}}{\delta_i}\right)^2}}.
\end{equation}

\paragraph{Cusum}
\cite{Kim_etal2014}
The range of magnitude cumulative sums:
\begin{equation}
    \max_j(S_j) - \min_j(S_j),
\end{equation}
where
\begin{equation}
    S_j \equiv \frac1{N\,\sigma_m} \sum_{i=0}^j{\left(m_i - \mean\right)}.
\end{equation}

\paragraph{Inter Percentile Range}
$\displaystyle Q(1-p)-Q(p),$ where $Q(n)$ and $Q(d)-$ $n$-th and $d$-th quantile of magnitude sample. We use $p = 0.25$ in our analysis.

\paragraph{Percent Amplitude}
Maximum deviation of magnitude from its median:
$\displaystyle \max_{i} |m_{i} \ -\ \text{Median}( m)|$

\paragraph{Maximum Slope}
Maximum slope between two sub-sequential observations:
$\displaystyle \max_{i=0\dotsc N-2}\left| \frac{m_{i+1} -m_{i}}{t_{i+1} -t_{i}}\right|$

\paragraph{Kurtosis}
The kurtosis of the magnitude distribution
\begin{equation}
    G_2 \equiv \frac{N\,(N + 1)}{(N - 1)(N - 2)(N - 3)} \frac{\sum_i(m_i - \mean)^4}{\sigma_m^4} - \frac{3(N+1)^2}{(N-2)(N-3)}.
\end{equation}

\paragraph{Mean Variance}
Standard deviation to mean ratio:
$\displaystyle \frac{\sigma _{m}}{\langle m\rangle }$

\paragraph{Otsu Split}~\cite{Otsu, Plateau_2023}
Difference of subset means, standard deviation of the lower subset, standard deviation of the upper
subset and lower-to-all observation count ratio for two subsets of magnitudes obtained by Otsu's method split.
Otsu's method is used to perform automatic thresholding. The algorithm returns a single threshold that separate values into two classes. This threshold is determined by minimizing intra-class intensity variance $\sigma^2_{W}=w_0\sigma^2_0+w_1\sigma^2_1$, or equivalently, by maximizing inter-class variance $\sigma^2_{B}=w_0 w_1 (\mu_1-\mu_0)^2$. There can be more than one extremum. In this case, the algorithm returns the minimum threshold.


\bsp	
\label{lastpage}
\end{document}